\newcommand{\gweak}{g_{\text{weak}}}
\newcommand{\mweak}{m_{\text{weak}}}

\newcommand{\gev}{\text{GeV}}

\newcommand{\cm}{\text{cm}}

\newcommand{\eg}{{\em e.g.}}

\newcommand{\eqref}[1]{Eq.~(\ref{#1})}

\newcommand{\figref}[1]{Fig.~\ref{fig:#1}}

\newcommand{\sigmaSI}{\sigma_{\text{SI}}}

\newcommand{\text}[1]{{\rm #1}}
\newcommand{\rem}[1]{{}}

\documentclass[
    ,final            
  ]
  {aipproc}

\layoutstyle{6x9}

\begin{document}

\title{Dark Matter Phenomenology\footnote{To appear in the 
Proceedings of the Tenth Conference on the Intersections of Particle
and Nuclear Physics (CIPANP 2009), San Diego, California, 26-31 May
2009.}}

\classification{95.35.+d \vspace*{-.06in} }


\author{Jonathan L.~Feng}{
  address={Department of Physics and Astronomy, University of
California, Irvine, California 92697, USA} }

\begin{abstract}
I review recent developments in the direct and indirect detection of
dark matter and new candidates beyond the WIMP paradigm.
\end{abstract}

\maketitle


\section{WIMP Dark Matter}
\vspace*{-.06in}

In recent years, the amount of dark matter in the Universe has become
precisely known, but its particle identity remains a mystery. Current
observational constraints require that the bulk of dark matter be
non-baryonic, cold or warm, and stable or long-lived.  These
constraints are easy to satisfy, and viable candidates have been
proposed with masses and interaction strengths that span many, many
orders of magnitude.

At the same time, there are strong reasons to focus on candidates with
masses around the weak scale $\mweak \sim 100~\gev$.  Despite
significant progress since this scale was first identified in the work
of Fermi in the 1930's, the origin of $\mweak$ is still unknown, and
every attempt to understand it so far introduces new particles at this
scale.  

Furthermore, the relic density of such particles naturally reproduces
the required dark matter density.  If a new (heavy) particle $X$ is
initially in thermal equilibrium, it can be shown that its relic
density today is
\begin{equation}
\Omega_X \propto \frac{1}{\langle \sigma_{\text{ann}} v \rangle}
\sim \frac{m_X^2}{g_X^4} \ ,
\label{omega}
\end{equation}
where $\langle \sigma_{\text{ann}} v \rangle$ is the
thermally-averaged annihilation cross section, and we have
parametrized it in terms of a mass scale $m_X$ and coupling constant
$g_X$ that characterize $XX$ annihilation.  Including all relevant
numerical factors, for weakly-interacting massive particles (WIMPs)
with $m_X \sim \mweak$ and $g_X \sim \gweak \simeq 0.65$, the resulting
relic density is $\Omega_X \sim 0.1$, near the required value for dark
matter $\Omega_{\text{DM}} \simeq 0.23$.  This remarkable coincidence
is the ``WIMP miracle.''  It implies that particle physics theories
designed to explain the origin of the weak scale often naturally
contain particles with the right relic density to be dark matter.

For WIMPs $X$ to have the right relic density, they must annihilate
through $XXff$ interactions, where, in the simplest cases, $f$ denotes
any of the known standard model particles.  This implies that dark
matter can be detected through $Xf \to Xf$ scattering (direct
detection), or through its annihilations $XX \to f \bar{f}$ (indirect
detection).  Both strategies are currently vigorously pursued, and
some recent highlights are briefly reviewed in the next two sections.

\vspace*{-.06in}
\section{Direct Detection}
\vspace*{-.06in}

\begin{figure}
\includegraphics[height=.3\textheight]{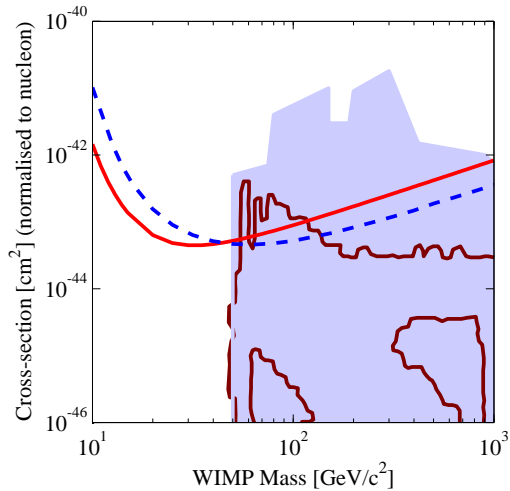}
\caption{Current bounds~\cite{Gaitskell} on spin-independent
WIMP-nucleon cross sections $\sigmaSI$ from
XENON10~\cite{Angle:2007uj} (solid red) and CDMS~\cite{Ahmed:2008eu}
(dashed blue), along with predictions from the general minimal
supersymmetric standard model (MSSM)~\cite{Baltz:2002ei} (shaded) and
minimal supergravity~\cite{Baltz:2004aw} (outlined).
\label{fig:direct}}
\end{figure}

Dark matter scattering is either spin-independent or
spin-dependent~\cite{Goodman:1984dc}.  Current bounds and
supersymmetric predictions for spin-independent scattering off
nucleons are shown in \figref{direct}.  As can be seen, current bounds
do not test the bulk of supersymmetric parameter space. The
experiments are improving rapidly, however, and in the coming year,
sensitivities to cross sections of $\sigmaSI \sim 10^{-45}~\cm^2$ are
possible.

How significant will this progress be?  As evident in \figref{direct},
the full range of supersymmetric predictions will not be probed for
the foreseeable future.  However, many well-known supersymmetric
theories predict $\sigmaSI \sim 10^{-44}~\cm^2$.  Supersymmetric
theories suffer from flavor and CP problems: the introduction of
squarks and sleptons with generic flavor mixing and weak scale masses
induces contributions to $K-\bar{K}$ mixing, $\mu \to e \gamma$, the
electric dipole moments of the neutron and electron, and a host of
other flavor- or CP-violating observables that badly violate known
constraints.  One generic solution to this problem is to assume heavy
squarks and sleptons, so that they decouple and do not affect
low-energy observables.

In general, the dominant contributions to neutralino annihilation are
$\chi \chi \to q \bar{q}, l \bar{l}$ through $t$-channel squarks and
sleptons, and $\chi \chi \to W^+ W^-, Z Z$ through $t$-channel
charginos and neutralinos.  In these decoupling theories, the first
diagrams are ineffective, and so annihilation takes place through the
second class of diagrams.  Essentially, two parameters enter these
diagrams: the neutralino's mass and its Higgsino content.  To keep the
relic density constant, larger $\chi$ masses are compensated by larger
Higgsino components.  In these models, then, the supersymmetry
parameter space is greatly reduced, with $\sigmaSI$ essentially
determined by the $\chi$ mass.  More detailed study shows that
$\sigmaSI$ is in fact fairly constant, with values near
$10^{-44}~\cm^2$, irrespective of mass.

In the next year or so, then, direct detection will probe many
well-known supersymmetric models with widely varying motivations, from
focus point models~\cite{Feng:2000gh} to split
supersymmetry~\cite{Pierce:2004mk}.  So far, direct detection
experiments have trimmed a few fingernails off the body of
supersymmetry parameter space, but if nothing is seen in the coming
few years, it is arms and legs that will have been lopped off.

In addition to the limits described above, the DAMA experiment
continues to find a signal in annual modulation~\cite{Drukier:1986tm}
with period and maximum at the expected values~\cite{Bernabei:2008yi}.
{}From a theorist's viewpoint, the DAMA/LIBRA result has been puzzling
because the signal, if interpreted as spin-independent elastic
scattering, seemingly implied dark matter masses and scattering cross
sections that have been excluded by other experiments.  Inelastic
scattering has been put forward as one
solution~\cite{TuckerSmith:2001hy}.  More recently,
astrophysics~\cite{Gondolo:2005hh} and
channeling~\cite{Drobyshevski:2007zj,Bernabei:2007hw}, a condensed
matter effect that effectively lowers the threshold for crystalline
detectors, have been proposed as possible remedies to allow elastic
scattering to explain DAMA without violating other constraints. If
these indications are correct, the favored parameters are $m_X \sim
5~\gev$ and $\sigmaSI \sim 10^{-39}~\cm^2$.  This mass is lower than
typically expected, but even massless neutralinos are allowed if one
relaxes the constraint of gaugino mass
unification~\cite{Dreiner:2009ic}.  The cross section is, however,
very large; it may be achieved in corners of MSSM parameter
space~\cite{Bottino:2007qg}, but is more easily explained in
completely different frameworks, such as those discussed below.

\vspace*{-.06in} 
\section{Indirect Detection}
\vspace*{-.06in}

Indirect searches look for particles produced when dark matter
particles decay or pair annihilate.  In contrast to direct detection,
there have been many reported anomalies in indirect detection, which
have been interpreted as possible evidence for dark matter.

\begin{figure}
\includegraphics[height=.25\textheight]{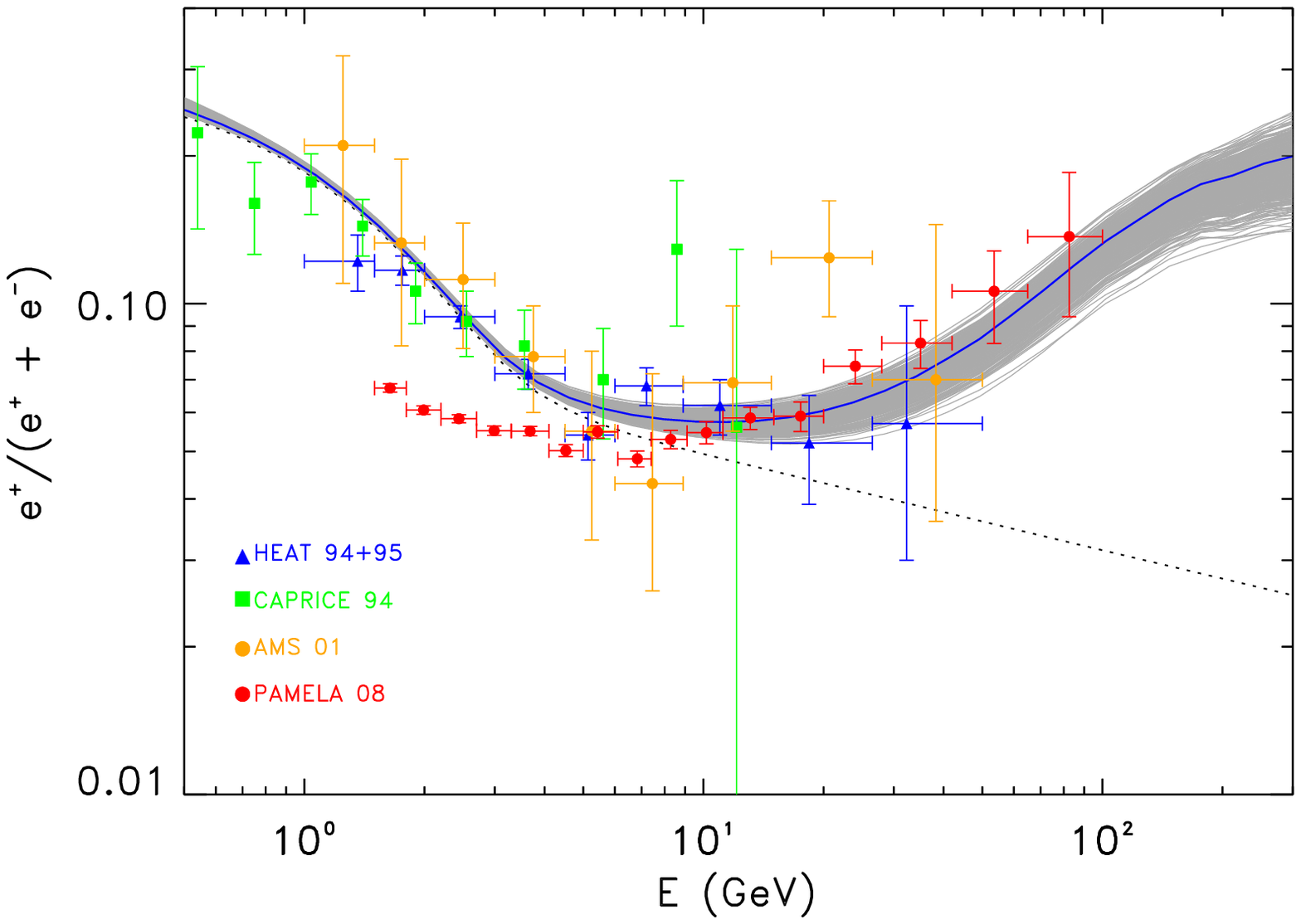}
\includegraphics[height=.25\textheight]{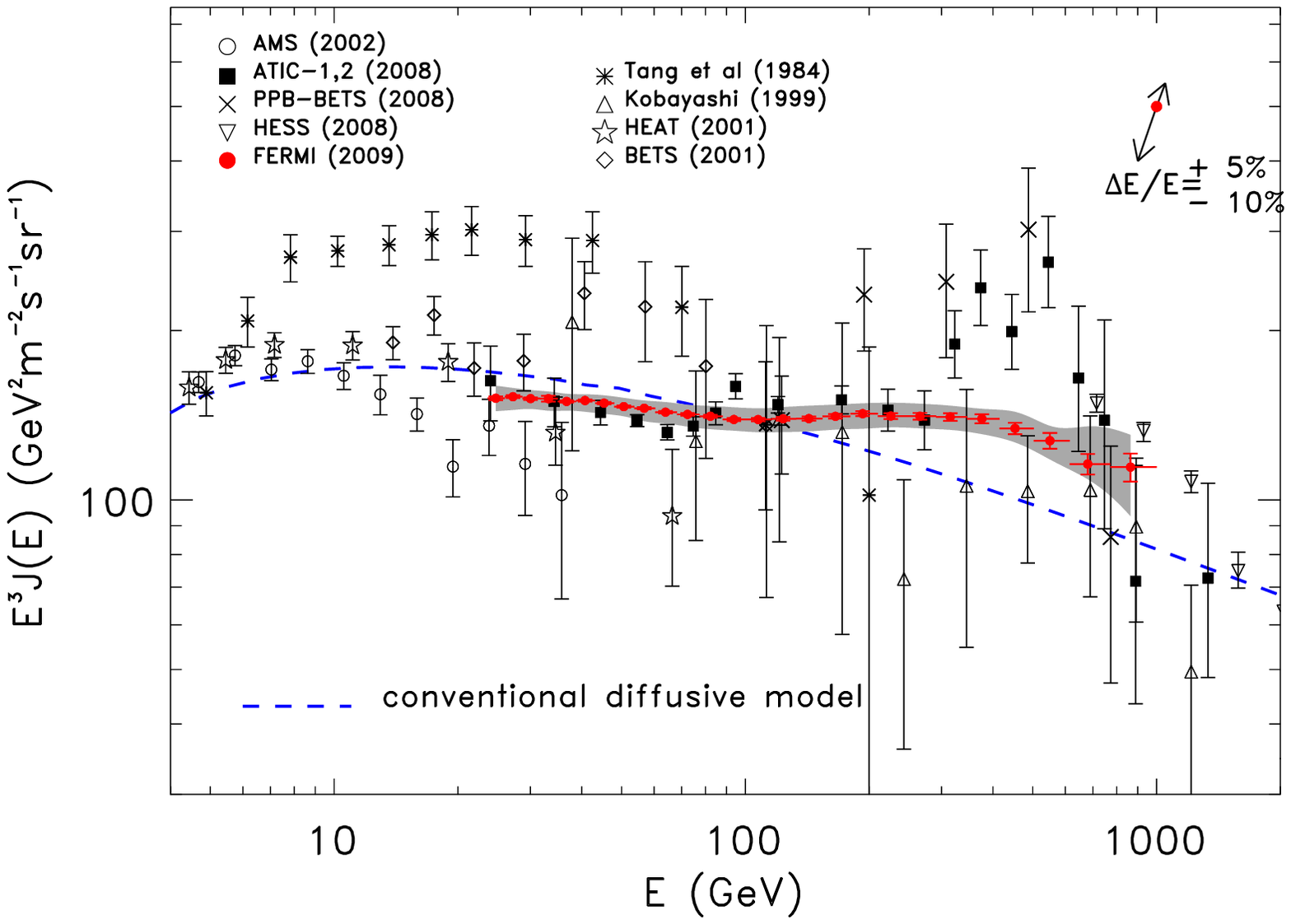}
\caption{The positron fraction measurements and predictions of
pulsars with various parameters~\cite{Grasso:2009ma} (left) and the
total $e^+ +e^-$ flux measured by ATIC, Fermi, and other
experiments~\cite{Abdo:2009zk} (right).
\label{fig:fermi}}
\end{figure}

Recently, the PAMELA and ATIC Collaborations have measured the
positron fraction $e^+/(e^+ + e^-)$~\cite{Adriani:2008zr} and the
total $e^+ + e^-$ flux~\cite{:2008zzr}, respectively, at energies in
the range of 10 GeV to 1 TeV.  These measurements revealed excesses
above the expected cosmic ray background from
GALPROP~\cite{Strong:2009xj}, which have been interpreted as dark
matter signals. The PAMELA results are, however, consistent with
astrophysical sources, such as the predicted fluxes from pulsars, as
derived both long
before~\cite{1989ApJ...342..807B,Chi:1995id,2001A&A...368.1063Z} and
after~\cite{Hooper:2008kg,Yuksel:2008rf,Profumo:2008ms,Malyshev:2009tw,Grasso:2009ma}
the release of the PAMELA data. Results from a recent study scanning
over pulsar characteristics are given in \figref{fermi}.  The ATIC
results have now been supplemented by high statistics data from the
Fermi LAT Collaboration~\cite{Abdo:2009zk}, which sees no evidence for
a bump (see \figref{fermi}). Additional data on both cosmic rays and
gamma rays will provide further insights (\eg, recent gamma ray data
already further disfavor some dark matter
proposals~\cite{Profumo:2009uf}) and are eagerly anticipated from
these experiments and many others.

\vspace*{-.2in} 

\vspace*{-.06in}
\section{Hidden Dark Matter}
\vspace*{-.06in}

The DAMA and other anomalies are not easy to explain with canonical
WIMPs.  This has motivated new candidates.  All solid evidence for
dark matter is gravitational, and there is also strong evidence
against dark matter having strong or electromagnetic interactions.  A
logical alternative to the WIMP paradigm, then, is hidden dark matter,
that is, dark matter that has no standard model gauge interactions.
Hidden dark matter has been explored for decades~\cite{Kobsarev:1966}.
By considering this possibility, though, one seemingly loses (1) a
connection to central problems in particle physics, such as the
problem of electroweak symmetry breaking, (2) the WIMP miracle, and
(3) the non-gravitational signals discussed above, which are most
likely required if we are to identify dark matter.

\begin{figure}
\includegraphics[height=.2\textheight,clip=]{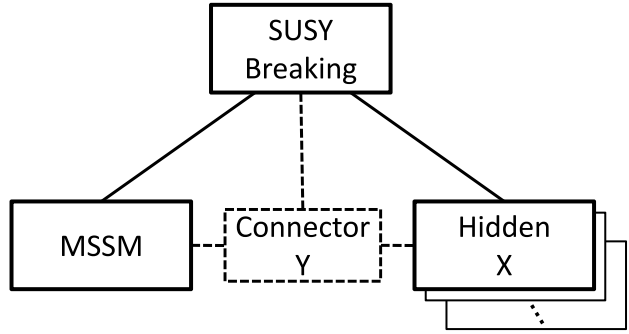}
\caption{Sectors in a supersymmetric hidden sector model.
Supersymmetry breaking is mediated to the MSSM and one or more hidden
sectors, which contain the dark matter particle $X$.  An optional
connector sector contains fields $Y$, charged under both MSSM and
hidden sector gauge groups, which induce signals in direct and
indirect searches and at colliders. {}From Ref.~\cite{Feng:2008ya}.
\label{fig:sectors}}
\end{figure}

In fact, however, hidden dark matter may have all three of the virtues
listed above.  Consider, for example, supersymmetric theories with
gauge-mediated supersymmetry breaking
(GMSB)~\cite{Dine:1994vc,Dine:1995ag}.  These models preserve the many
virtues of supersymmetry, while elegantly solving the flavor and CP
problems mentioned above.  Although minimal GMSB models contain only a
supersymmetry-breaking sector and the MSSM, in models that arise from
string theory, hidden sectors are ubiquitous.  As a concrete example,
we consider GMSB with an additional hidden sector, as depicted in
\figref{sectors}.

In these models, supersymmetry is broken when a chiral field in the
supersymmetry breaking sector gets a vacuum expectation value $\langle
S \rangle = M + \theta^2 F$.  The resulting superpartner masses in the
MSSM and hidden sectors are $m \sim \frac{g^2}{16 \pi^2} \frac{F}{M}$
and $m_X \sim \frac{g_X^2}{16 \pi^2} \frac{F}{M}$, and so
\begin{equation}
\frac{m_X}{g_X^2} \sim \frac{m}{g^2} \sim
\frac{F}{16 \pi^2 M} \ ;
\label{mxgx}
\end{equation}
that is, $m_X/g_X^2$ is determined solely by the
supersymmetry-breaking sector.  As this is exactly the combination of
parameters that determines the thermal relic density of \eqref{omega},
the hidden sector automatically includes a dark matter candidate that
has the desired thermal relic density, irrespective of its mass. This
is the ``WIMPless miracle'' --- in these models, a hidden sector
particle naturally has the desired thermal relic density, but it has
neither a weak-scale mass nor weak force
interactions~\cite{Feng:2008ya,Feng:2008mu}.

The WIMPless framework opens up many new possibilities for dark matter
signals without sacrificing the main motivations for WIMPs.  For
example, if there is an unbroken U(1) gauge symmetry in the hidden
sector, these dark matter particles self-interact through Rutherford
scattering, with a wide range of observable
implications~\cite{Ackerman:2008gi,Feng:2009mn}.

\begin{figure}
\includegraphics[height=.3\textheight]{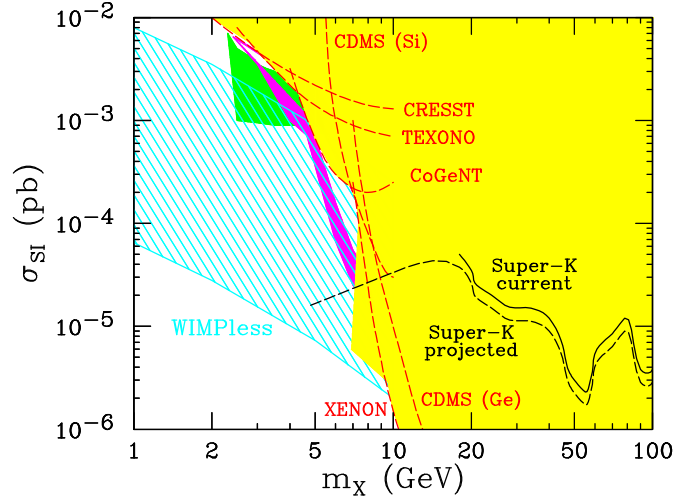}
\caption{Direct detection cross sections for spin-independent 
$X$-nucleon scattering as a function of dark matter mass $m_X$.  The
black lines are Super-K limits~\cite{Desai:2004pq} and projected
sensitivities. The magenta shaded region is DAMA-favored given
channeling and no streams~\cite{Petriello:2008jj}, and the medium
green shaded region is DAMA-favored at 3$\sigma$ given streams but no
channeling~\cite{Gondolo:2005hh}.  The light yellow shaded region is
excluded by the direct detection experiments indicated.  The
predictions of WIMPless models (with connector mass $m_Y = 400~\gev$
and $0.3 < \lambda_b < 1$) lie in the blue shaded region. {}From
Ref.~\cite{Feng:2008qn}.
\label{fig:superkdirect}}
\end{figure}

Alternatively, if there are connector particles, WIMPless dark matter
can explain the DAMA/LIBRA signal~\cite{Feng:2008dz}.  Suppose the
dark matter particle $X$ couples to standard model fermions $f$ via
exchange of a connector particle $Y$ through interactions ${\cal L} =
\lambda_f X \bar{Y}_L f_L + \lambda_f X \bar{Y}_R f_R$.  The Yukawa
couplings $\lambda_f$ are model-dependent.  If the dark matter couples
mainly to 3rd generation quarks, the dominant nuclear coupling of
WIMPless dark matter is to gluons via a loop of $b$-quarks.  The
predictions of this WIMPless dark matter model are given in
\figref{superkdirect}. In this WIMPless scenario, the small mass
required by DAMA is as natural as any other because of the WIMPless
miracle.  The large cross section is also easily achieved, because the
heavy $Y$ propagator flips chirality without Yukawa suppression, in
contrast to the case of neutralinos.  This explanation is testable
through searches for exotic 4th generation quarks at the Tevatron and
LHC.


\vspace*{-.06in} 
\begin{theacknowledgments}
\vspace*{-.06in} 
I thank my collaborators for their contributions to the work discussed
here, and Marvin Marshak and the organizers of CIPANP 2009 for a
beautifully organized conference.  This work was supported in part by
NSF grant PHY--0653656.
\end{theacknowledgments}

\vspace*{-.06in} 
\hyphenation{Post-Script Sprin-ger}


\begin{thebibliography}{40}
\expandafter\ifx\csname natexlab\endcsname\relax\def\natexlab#1{#1}\fi
\providecommand{\enquote}[1]{``#1''}
\expandafter\ifx\csname url\endcsname\relax
  \def\url#1{\texttt{#1}}\fi
\expandafter\ifx\csname urlprefix\endcsname\relax\def\urlprefix{URL }\fi
\providecommand{\eprint}[2][]{\url{#2}}

\bibitem[Goodman and Witten(1985)]{Goodman:1984dc}
M.~W. Goodman, and E.~Witten, \emph{Phys. Rev.} \textbf{D31}, 3059 (1985).

\bibitem[Gaitskell et~al.(2009)]{Gaitskell}
R.~Gaitskell, V.~Mandic, and J.~Filipini  (2009),
  \eprint{http://dmtools.berkeley.edu}.

\bibitem[Angle et~al.(2008)]{Angle:2007uj}
J.~Angle, et~al., \emph{Phys. Rev. Lett.} \textbf{100}, 021303 (2008),
  \eprint{0706.0039}.

\bibitem[Ahmed et~al.(2009)]{Ahmed:2008eu}
Z.~Ahmed, et~al., \emph{Phys. Rev. Lett.} \textbf{102}, 011301 (2009),
  \eprint{0802.3530}.

\bibitem[Baltz and Gondolo(2003)]{Baltz:2002ei}
E.~A. Baltz, and P.~Gondolo, \emph{Phys. Rev.} \textbf{D67}, 063503 (2003),
  \eprint{astro-ph/0207673}.

\bibitem[Baltz and Gondolo(2004)]{Baltz:2004aw}
E.~A. Baltz, and P.~Gondolo, \emph{JHEP} \textbf{10}, 052 (2004),
  \eprint{hep-ph/0407039}.

\bibitem[Feng et~al.(2000)]{Feng:2000gh}
J.~L. Feng, K.~T. Matchev, and F.~Wilczek, \emph{Phys. Lett.} \textbf{B482},
  388--399 (2000), \eprint{hep-ph/0004043}.

\bibitem[Pierce(2004)]{Pierce:2004mk}
A.~Pierce, \emph{Phys. Rev.} \textbf{D70}, 075006 (2004),
  \eprint{hep-ph/0406144}.

\bibitem[Drukier et~al.(1986)]{Drukier:1986tm}
A.~K. Drukier, K.~Freese, and D.~N. Spergel, \emph{Phys. Rev.} \textbf{D33},
  3495--3508 (1986).

\bibitem[Bernabei et~al.(2008{\natexlab{a}})]{Bernabei:2008yi}
R.~Bernabei, et~al., \emph{Eur. Phys. J.} \textbf{C56}, 333--355
  (2008{\natexlab{a}}), \eprint{0804.2741}.

\bibitem[Tucker-Smith and Weiner(2001)]{TuckerSmith:2001hy}
D.~Tucker-Smith, and N.~Weiner, \emph{Phys. Rev.} \textbf{D64}, 043502 (2001),
  \eprint{hep-ph/0101138}.

\bibitem[Gondolo and Gelmini(2005)]{Gondolo:2005hh}
P.~Gondolo, and G.~Gelmini, \emph{Phys. Rev.} \textbf{D71}, 123520 (2005),
  \eprint{hep-ph/0504010}.

\bibitem[Drobyshevski(2008)]{Drobyshevski:2007zj}
E.~M. Drobyshevski, \emph{Mod. Phys. Lett.} \textbf{A23}, 3077--3085 (2008),
  \eprint{0706.3095}.

\bibitem[Bernabei et~al.(2008{\natexlab{b}})]{Bernabei:2007hw}
R.~Bernabei, et~al., \emph{Eur. Phys. J.} \textbf{C53}, 205--213
  (2008{\natexlab{b}}), \eprint{0710.0288}.

\bibitem[Dreiner et~al.(2009)]{Dreiner:2009ic}
H.~K. Dreiner, et~al., \emph{Eur. Phys. J.} \textbf{C62}, 547--572 (2009),
  \eprint{0901.3485}.

\bibitem[Bottino et~al.(2008)]{Bottino:2007qg}
A.~Bottino, F.~Donato, N.~Fornengo, and S.~Scopel, \emph{Phys. Rev.}
  \textbf{D77}, 015002 (2008), \eprint{0710.0553}.

\bibitem[Adriani et~al.(2009)]{Adriani:2008zr}
O.~Adriani, et~al., \emph{Nature} \textbf{458}, 607--609 (2009),
  \eprint{0810.4995}.

\bibitem[Chang et~al.(2008)]{:2008zzr}
J.~Chang, et~al., \emph{Nature} \textbf{456}, 362--365 (2008).

\bibitem[Strong et~al.(2009)]{Strong:2009xj}
A.~W. Strong, et~al.  (2009), \eprint{0907.0559}.

\bibitem[{Boulares}(1989)]{1989ApJ...342..807B}
A.~{Boulares}, \emph{Astrophys.\ J.} \textbf{342}, 807--813 (1989).

\bibitem[Chi et~al.(1995)]{Chi:1995id}
X.~Chi, E.~C.~M. Young, and K.~S. Cheng, \emph{Astrophys. J.} \textbf{459},
  L83--L86 (1995).

\bibitem[{Zhang} and {Cheng}(2001)]{2001A&A...368.1063Z}
L.~{Zhang}, and K.~S. {Cheng}, \emph{Astron. Astrophys.} \textbf{368},
  1063--1070 (2001).

\bibitem[Hooper et~al.(2009)]{Hooper:2008kg}
D.~Hooper, P.~Blasi, and P.~D. Serpico, \emph{JCAP} \textbf{0901}, 025 (2009),
  \eprint{0810.1527}.

\bibitem[Yuksel et~al.(2009)]{Yuksel:2008rf}
H.~Yuksel, M.~D. Kistler, and T.~Stanev, \emph{Phys. Rev. Lett.} \textbf{103},
  051101 (2009), \eprint{0810.2784}.

\bibitem[Profumo(2008)]{Profumo:2008ms}
S.~Profumo  (2008), \eprint{0812.4457}.

\bibitem[Malyshev et~al.(2009)]{Malyshev:2009tw}
D.~Malyshev, I.~Cholis, and J.~Gelfand  (2009), \eprint{0903.1310}.

\bibitem[Grasso et~al.(2009)]{Grasso:2009ma}
D.~Grasso, et~al.  (2009), \eprint{0905.0636}.

\bibitem[Abdo et~al.(2009)]{Abdo:2009zk}
A.~A. Abdo, et~al., \emph{Phys. Rev. Lett.} \textbf{102}, 181101 (2009),
  \eprint{0905.0025}.

\bibitem[Profumo and Jeltema(2009)]{Profumo:2009uf}
S.~Profumo, and T.~E. Jeltema, \emph{JCAP} \textbf{0907}, 020 (2009),
  \eprint{0906.0001}.

\bibitem[Kobsarev et~al.(1966)]{Kobsarev:1966}
I.~Y. Kobsarev, L.~B. Okun, and I.~Y. Pomeranchuk, \emph{Sov.\ J.\ Nucl.\
  Phys.} \textbf{3}, 837 (1966).

\bibitem[Dine et~al.(1995)]{Dine:1994vc}
M.~Dine, A.~E. Nelson, and Y.~Shirman, \emph{Phys. Rev.} \textbf{D51},
  1362--1370 (1995), \eprint{hep-ph/9408384}.

\bibitem[Dine et~al.(1996)]{Dine:1995ag}
M.~Dine, A.~E. Nelson, Y.~Nir, and Y.~Shirman, \emph{Phys. Rev.} \textbf{D53},
  2658--2669 (1996), \eprint{hep-ph/9507378}.

\bibitem[Feng and Kumar(2008)]{Feng:2008ya}
J.~L. Feng, and J.~Kumar, \emph{Phys. Rev. Lett.} \textbf{101}, 231301 (2008),
  \eprint{0803.4196}.

\bibitem[Feng et~al.(2008{\natexlab{a}})]{Feng:2008mu}
J.~L. Feng, H.~Tu, and H.-B. Yu, \emph{JCAP} \textbf{0810}, 043
  (2008{\natexlab{a}}), \eprint{0808.2318}.

\bibitem[Ackerman et~al.(2009)]{Ackerman:2008gi}
L.~Ackerman, M.~R. Buckley, S.~M. Carroll, and M.~Kamionkowski, \emph{Phys.
  Rev.} \textbf{D79}, 023519 (2009), \eprint{0810.5126}.

\bibitem[Feng et~al.(2009{\natexlab{a}})]{Feng:2009mn}
J.~L. Feng, M.~Kaplinghat, H.~Tu, and H.-B. Yu, \emph{JCAP} \textbf{0907}, 004
  (2009{\natexlab{a}}), \eprint{0905.3039}.

\bibitem[Feng et~al.(2008{\natexlab{b}})]{Feng:2008dz}
J.~L. Feng, J.~Kumar, and L.~E. Strigari, \emph{Phys. Lett.} \textbf{B670},
  37--40 (2008{\natexlab{b}}), \eprint{0806.3746}.

\bibitem[Desai et~al.(2004)]{Desai:2004pq}
S.~Desai, et~al., \emph{Phys. Rev.} \textbf{D70}, 083523 (2004),
  \eprint{hep-ex/0404025}.

\bibitem[Petriello and Zurek(2008)]{Petriello:2008jj}
F.~Petriello, and K.~M. Zurek, \emph{JHEP} \textbf{09}, 047 (2008),
  \eprint{0806.3989}.

\bibitem[Feng et~al.(2009{\natexlab{b}})]{Feng:2008qn}
J.~L. Feng, J.~Kumar, J.~Learned, and L.~E. Strigari, \emph{JCAP}
  \textbf{0901}, 032 (2009{\natexlab{b}}), \eprint{0808.4151}.

\end{thebibliography}
\end{document}